%% file: main.tex
\documentclass[11pt]{article}

\usepackage[preprint]{acl}

\usepackage{times}
\usepackage{latexsym}
\usepackage[T1]{fontenc}
\usepackage[utf8]{inputenc}
\usepackage{microtype}
\usepackage{inconsolata}
\usepackage{graphicx}

\usepackage{booktabs}
\usepackage{multirow}
\usepackage{amsmath}
\usepackage{amssymb}
\usepackage{tikz}
\usepackage{enumitem}
\usetikzlibrary{positioning,arrows.meta,shapes,backgrounds,fit,calc}

\title{ProRetrieval: Learning to Orchestrate Hybrid Search\\
via Executable Program Synthesis}

\author{
  Chengsong You\textsuperscript{1,3},\;
  Zhen Sun\textsuperscript{1,3},\;
  Yunhai Hu\textsuperscript{4},\;
  Junwei Zhou\textsuperscript{2}\thanks{Corresponding authors.},\;
  Xiaoyu Cao\textsuperscript{1},\;
  Binyu Li\textsuperscript{1},\;
  Ziyan Zhao\textsuperscript{1},\\
  \bfseries Weiyao Wang\textsuperscript{1},\;
  Liren Lu\textsuperscript{3},\;
  Zhijie Ye\textsuperscript{3},\;
  Yumo Cao\textsuperscript{1},\;
  Yitao Long\textsuperscript{4},\;
  Yiwei Xu\textsuperscript{1,3},\;
  Qiyi Jiang\textsuperscript{1},\\
  \bfseries Xuanyi Fu\textsuperscript{5},\;
  Yufan Chen\textsuperscript{5},\;
  Yilun Li\textsuperscript{5},\;
  Rongkang Xiong\textsuperscript{5},\;
  Yiran Zou\textsuperscript{1},\;
  Nan Du\textsuperscript{2}$^*$ \\[6pt]
  \textsuperscript{1}Thin Red Line \quad
  \textsuperscript{2}Matter Innovation Inc. \quad
  \textsuperscript{3}East China Normal University \\
  \textsuperscript{4}New York University \quad
  \textsuperscript{5}Independent Researcher \\[3pt]
  {\small \texttt{51275901122@stu.ecnu.edu.cn} \quad
  \texttt{sunzhen@thinredline.com.cn} \quad
  \texttt{yh5961@nyu.edu}} \\
  {\small \texttt{zjw330501@gmail.com} \quad
  \texttt{frankdu@matter.ai}}
}

\begin{document}
\maketitle

\input{sections/abstract}
\input{sections/introduction}
\input{sections/related_work}
\input{sections/method}
\input{sections/experiments}
\input{sections/conclusion}

\section*{Limitations}
\input{sections/limitations}

\section*{Ethics Statement}
ProRetrieval is a retrieval orchestrator that synthesizes executable programs to locate existing documents; it does not generate user-facing content and therefore poses minimal risk of producing harmful or misleading text.
All experiments use publicly available research datasets: Amazon ESCI (product metadata) and Enron (public email corpus released for research).
We do not attempt to re-identify individuals in the Enron dataset and use only metadata and text fields for retrieval evaluation.
We note that structured filtering conditions (e.g., brand or price constraints) could in some applications systematically exclude certain groups or information sources, and that biases in the underlying embedding models may propagate into retrieval rankings; practitioners should audit filter distributions and embedding fairness when deploying program-based retrieval in sensitive domains.

\bibliography{references}

\appendix
\input{sections/appendix}

\end{document}

%% file: sections/abstract.tex
\begin{abstract}
Real-world retrieval often composes structured constraints with semantic intents over text and images through arbitrary Boolean logic. Existing hybrid pipelines such as reciprocal rank fusion or self-querying retrievers admit only a fixed form of composition, while recent reinforcement-learning retrievers train the language model as a \emph{query generator} for a single backend, leaving the orchestration of heterogeneous retrieval paths outside its action space. We propose \textbf{ProRetrieval}, which recasts the language model as a \emph{retrieval orchestrator}: given a natural-language query, it synthesizes an executable program in a hybrid DSL interleaving SQL operators over structured fields with vector-retrieval primitives over text and images, with SQL itself providing the logical algebra that fuses heterogeneous candidate sets. We train Qwen3-4B with GRPO and DAPO under a hierarchical four-term reward, and evaluate on two new benchmarks built from Amazon products and Enron email. Our 4B model surpasses GPT-5.5 (Hit@1 $0.81$ vs.\ $0.69$ on e-commerce; $0.91$ vs.\ $0.86$ on email) and Claude Opus 4.7 and a comprehensive suite of retrieval, LLM-augmented, structured-query, and graph-based baselines. Code: \url{https://anonymous.4open.science/r/ProRetrieval/}; data: \url{https://huggingface.co/datasets/anonymous-7219/ProRetrieval}.
\end{abstract}

%% file: sections/introduction.tex
\section{Introduction}
\label{sec:intro}

\input{sections/fig_framework}

Real-world retrieval rarely reduces to either symbolic filtering or pure semantic matching. A user may ask for \emph{``last week's emails from the marketing team about the Q3 budget, but not the archived ones''}, or for \emph{``a wireless headphone under \$200 from Sony or Bose, with strong noise-cancelling reviews, in matte black''}. Such queries share two properties: information is split across heterogeneous backends (relational fields, text and image indices), and constraints are composed by arbitrary Boolean connectives (disjunction, negation, and nested groups). Solving them requires not only understanding \emph{what} the user wants, but deciding \emph{how} to find it: which conditions to push down as exact filters, which intents to pose as semantic queries, and how to logically combine the resulting candidate sets.

Existing methods fall short along two axes. On \textbf{expressive power}, pure retrievers~\citep{robertson2009bm25,karpukhin2020dpr} cover a single modality; fusion pipelines~\citep{cormack2009rrf} express only weighted disjunction; and self-querying retrievers~\citep{langchain2023selfquery} admit simple key--value conjunctions but not disjunction, negation, or nested predicates. On \textbf{orchestration capability}, recent RL approaches such as Search-R1~\citep{jin2025searchr1} and DeepRetrieval~\citep{jiang2025deepretrieval} train models to issue better queries to a \emph{single} backend but cannot compose across heterogeneous stores. Hybrid retrieval therefore requires enlarging the model's role from \emph{query generator} to \emph{retrieval orchestrator}, with an action space that is itself an executable program over a logically complete operator set.

We propose \textbf{ProRetrieval}, a framework that trains a small language model to orchestrate hybrid retrieval by synthesizing executable retrieval programs. Our key observation is that SQL's logical algebra (selection, conjunction, disjunction, negation, and nested subqueries) can serve as the orchestration skeleton, while vector retrieval is incorporated as a first-class primitive whose results plug into SQL through candidate-set placeholders (e.g., \texttt{id IN <text\_0>}). This reduces orchestration to learnable program generation in a hybrid DSL while inheriting SQL's logical completeness for free. We train Qwen3-4B with GRPO~\citep{shao2024grpo} and DAPO~\citep{wang2025dapo} under a hierarchical four-term reward (format, execution, result, length) that mirrors the layered correctness requirements of executable program generation. To evaluate the framework we construct two hybrid retrieval benchmarks from public corpora: Amazon e-commerce (structured + text + image) and Enron email (structured + text), with queries spanning single-leaf to nested Boolean expressions.

On both benchmarks our 4B model surpasses GPT-5.5 (Hit@1 $0.81$ vs.\ $0.69$ on e-commerce; $0.91$ vs.\ $0.86$ on email) and a comprehensive suite of retrieval, reranking, structured-query, and graph-based baselines. In summary, our contributions are:

\begin{itemize}
\setlength{\itemsep}{2pt}
\setlength{\parskip}{0pt}
\item \textbf{A new paradigm: retrieval program synthesis.} We reformulate hybrid retrieval as program generation in a DSL that unifies SQL operators with multimodal vector primitives, lifting the model's action space from a textual query to a logically complete executable program. The expressiveness strictly subsumes that of fixed-fusion pipelines and single-backend RL retrievers.
\item \textbf{A hierarchical reward for executable program RL.} We introduce a four-term composite reward (format, execution, result, length) that mirrors the layered correctness requirements of program generation, generalizing the single-signal rewards used by prior RL retrievers.
\item \textbf{Two hybrid retrieval benchmarks and a comprehensive empirical study.} We release benchmarks over Amazon products and Enron email, together with four paradigms of baselines, and demonstrate that a 4B open-source model trained with our framework consistently outperforms frontier commercial LLMs at a fraction of their inference cost.
\end{itemize}

%% file: sections/fig_framework.tex
\begin{figure*}[t]
\centering
\includegraphics[width=\linewidth]{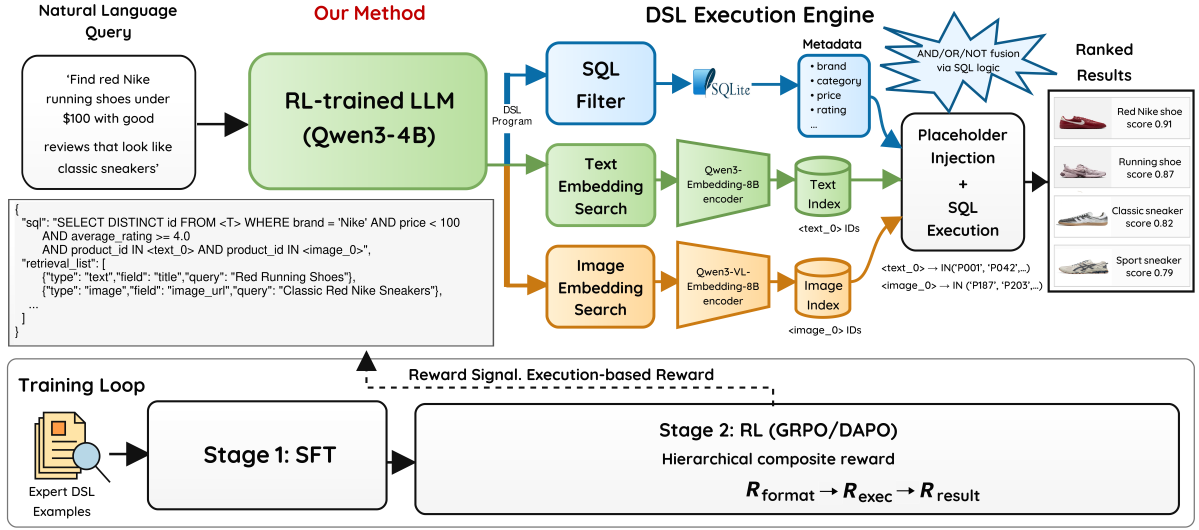}
\caption{Overview of \textbf{ProRetrieval}. Given a natural-language query, the RL-trained policy emits a hybrid DSL program that combines an SQL filter with text and image vector-retrieval primitives. The DSL execution engine dispatches each primitive to its embedding-based backend, substitutes the returned candidate sets into the SQL through placeholder injection, and lets the SQL engine perform the final AND/OR/NOT fusion to produce ranked results. The model is trained in two stages: SFT on expert DSL examples followed by RL (GRPO\,/\,DAPO) under a hierarchical composite reward over format, executability, and result quality.}
\label{fig:framework}
\end{figure*}

%% file: sections/related_work.tex
\section{Related Work}
\label{sec:related}

\paragraph{Hybrid retrieval.}
Sparse and dense retrievers~\citep{robertson2009bm25,karpukhin2020dpr,khattab2020colbert} and modern multimodal embedders~\citep{chen2024bgem3,faysse2024colpali}, optionally followed by cross-encoder rerankers~\citep{nogueira2019passage,xiao2023bge}, each operate within a single representational space. Industry pipelines combine them at the rank level via reciprocal rank fusion~\citep{cormack2009rrf}, which captures only a weighted disjunction, or through self-querying retrievers~\citep{langchain2023selfquery}, whose filter language admits only conjunctions of simple key--value predicates. Knowledge-graph retrievers such as GraphRAG~\citep{edge2024graphrag} introduce an orthogonal structured signal and are complementary to our field-centric design. Across this line the orchestration logic is fixed by the pipeline designer and cannot adapt per query.

\paragraph{RL for retrieval.}
Recent work uses RL to train language models that interact with retrieval: Search-R1~\citep{jin2025searchr1} interleaves reasoning with calls to a search engine; DeepRetrieval~\citep{jiang2025deepretrieval} optimizes structured Boolean / SQL-like query reformulation against a single backend; s3~\citep{jiang2025s3}, R1-Searcher~\citep{song2025r1searcher}, and follow-ups~\citep{chen2025research,li2025r3rag} explore further variants, and SQL-R1~\citep{sqlr12025} applies RL to text-to-SQL. These methods improve the queries a model issues, but the action space is restricted to the type of query a single fixed backend accepts and none orchestrates multiple heterogeneous backends with arbitrary logical composition. In particular, SQL-R1 cannot express semantic similarity queries (e.g., ``shoes that look like classic sneakers'') because SQL lacks vector retrieval operators; our action space ablation (\S\ref{sec:action_space}) quantifies this gap: SQL-only achieves 0.650 Hit@1 vs.\ 0.809 for the full hybrid DSL. We compare directly with Search-R1, DeepRetrieval, and SQL-R1.

\paragraph{DSLs for retrieval.}
DSLs have long bridged language models and symbolic computation in program synthesis~\citep{balog2017deepcoder,chen2021codex} and tool use~\citep{yao2023react,schick2023toolformer}. Closest to our work, SUQL~\citep{liu2024suql} extends SQL with a \texttt{SUMMARY} operator for conversational QA, and Text2VectorSQL~\citep{wang2025text2vectorsql} proposes a unified SQL interface with vector search predicates. Text2VectorSQL embeds vector operations as SQL UDFs, coupling retrieval and fusion in a database-native engine; our placeholder injection decouples them, enabling independent optimization and extension to image modality. The two systems target different tasks---Text2VectorSQL evaluates SQL execution accuracy on Spider/BIRD-style benchmarks, whereas we evaluate document retrieval quality (Hit@1, MRR)---and our SFT baseline already represents the same supervised-learning paradigm, upon which RL provides a further $+12.9$\,pp gain (\S\ref{sec:main_results}). Other approaches include TableQuery~\citep{tablequery2024} and CRUSH~\citep{crush2024}, focusing on structured-only retrieval without vector primitives. All of the above rely on prompting or SFT. ProRetrieval differs by treating the DSL as a learnable policy space optimized end-to-end by RL with a hierarchical reward tailored to executable hybrid programs.

%% file: sections/method.tex
\section{Method}
\label{sec:method}

ProRetrieval trains a small language model to act as a retrieval orchestrator that synthesizes executable hybrid retrieval programs. The framework, illustrated in Figure~\ref{fig:framework}, comprises three components: (i)~a \emph{hybrid retrieval DSL} that interleaves SQL operators with text and image vector-retrieval primitives (\S\ref{sec:dsl}); (ii)~a \emph{query construction pipeline} that automatically produces $(\text{NL query},~\text{gold DSL},~\text{ground-truth documents})$ triples covering single-leaf to nested Boolean expressions (\S\ref{sec:data}); and (iii)~a \emph{two-stage training procedure} that warm-starts the model with supervised fine-tuning on gold DSL and then optimizes it with reinforcement learning under a hierarchical composite reward (\S\ref{sec:train}).

\subsection{Hybrid Retrieval DSL}
\label{sec:dsl}

\paragraph{Syntax.}
A program in our DSL is a JSON object with two fields:
\begin{quote}
\small
\begin{verbatim}
{
  "sql": "SELECT DISTINCT id FROM <T>
          WHERE <conditions>",
  "retrieval_list": [
    {"type": "text",  "field": "<f1>",
     "query": "<semantic phrase>"},
    {"type": "image", "field": "<f2>",
     "query": "<visual phrase>"},
    ...
  ]
}
\end{verbatim}
\end{quote}
The \texttt{sql} field is a standard SQL query over the structured fields of the underlying table; it may invoke vector retrieval through placeholders of the form \texttt{<text\_$k$>} or \texttt{<image\_$k$>}, where $k$ indexes the corresponding entry in \texttt{retrieval\_list}. Each entry specifies the modality, the field on which similarity is computed, and the textual query to be embedded. Because the placeholders appear as candidate-set constraints inside arbitrary SQL clauses, the program inherits SQL's full logical algebra: \textsc{and}, \textsc{or}, \textsc{not}, and nested subqueries can freely combine structured predicates with semantic ones.

\paragraph{Execution.}
At execution time, each entry in \texttt{retrieval\_list} is dispatched to the corresponding backend (a text or image encoder followed by approximate nearest-neighbor search), which returns the top-$K$ candidate identifiers $\mathcal{C}_k$ (we use $K{=}20$ throughout all experiments). The placeholders are then substituted by the corresponding sets, transforming semantic similarity into set membership predicates (e.g., \texttt{id IN (P001, P042, ...)}). The resulting SQL is executed against the relational backend, which performs the final logical composition via its native query engine. This two-phase design (vector retrieval followed by relational fusion) decouples the embedding computation from the logical orchestration, allowing each component to be optimized independently.

\begin{figure*}[t]
\centering
\includegraphics[width=\textwidth]{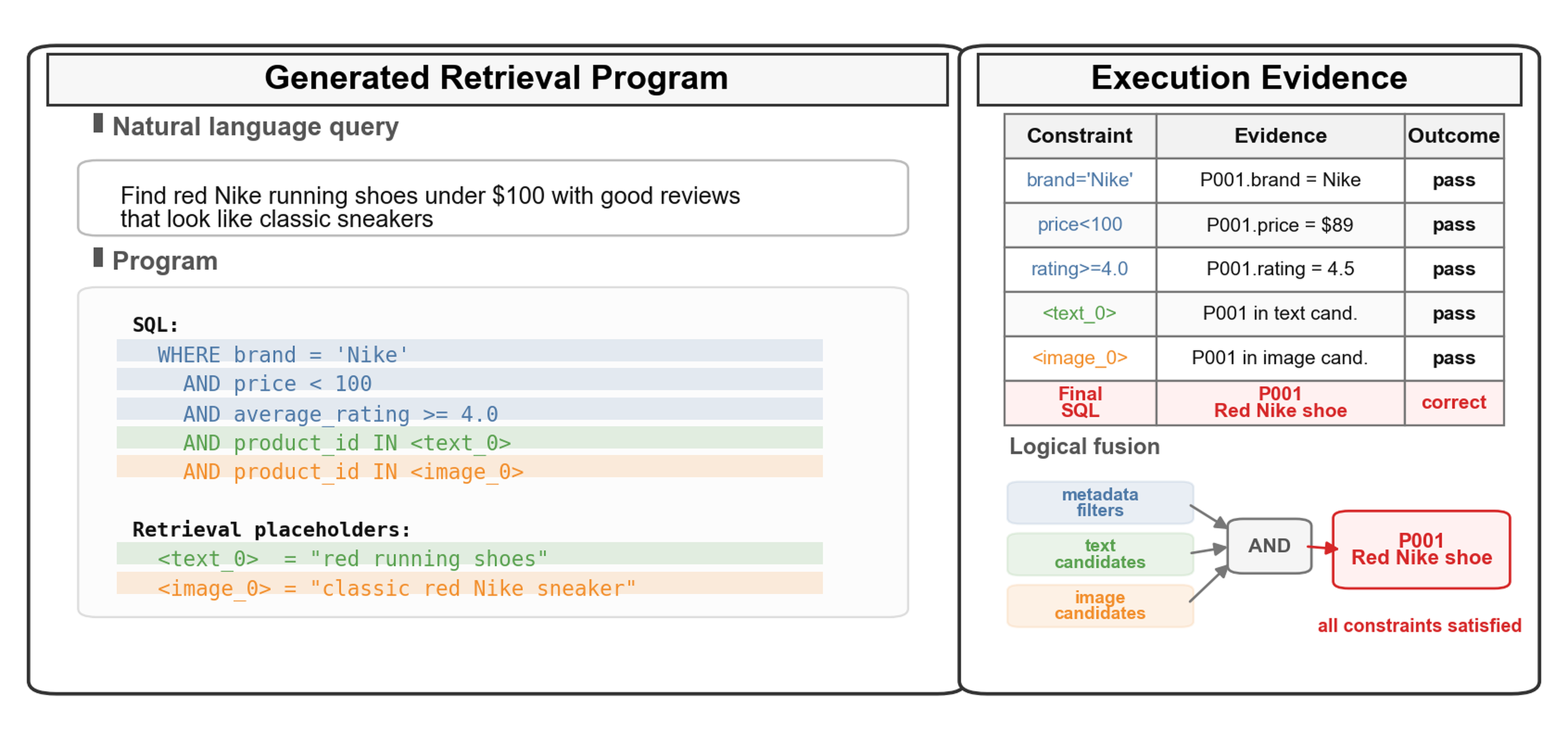}
\caption{Worked example of the hybrid DSL. The program issues an SQL filter together with text and image vector primitives; the execution engine substitutes their returned candidate sets into the SQL through placeholder injection and verifies each leaf against the retrieved gold document.}
\label{fig:case_study}
\end{figure*}

The DSL strictly subsumes the action spaces of prior methods: reciprocal rank fusion corresponds to a disjunction of vector predicates without structured filtering; self-querying retrievers to a single vector predicate conjoined with key--value filters; pure text-to-SQL to the empty \texttt{retrieval\_list}. This hierarchy means any query expressible by a simpler paradigm is expressible in our DSL, while the converse does not hold---queries requiring cross-modal logical composition (e.g., structured AND semantic AND visual) are unique to the full DSL. Figure~\ref{fig:case_study} shows a worked example.

\subsection{Query Construction Pipeline}
\label{sec:data}

\begin{figure}[t]
\centering
\includegraphics[width=\columnwidth]{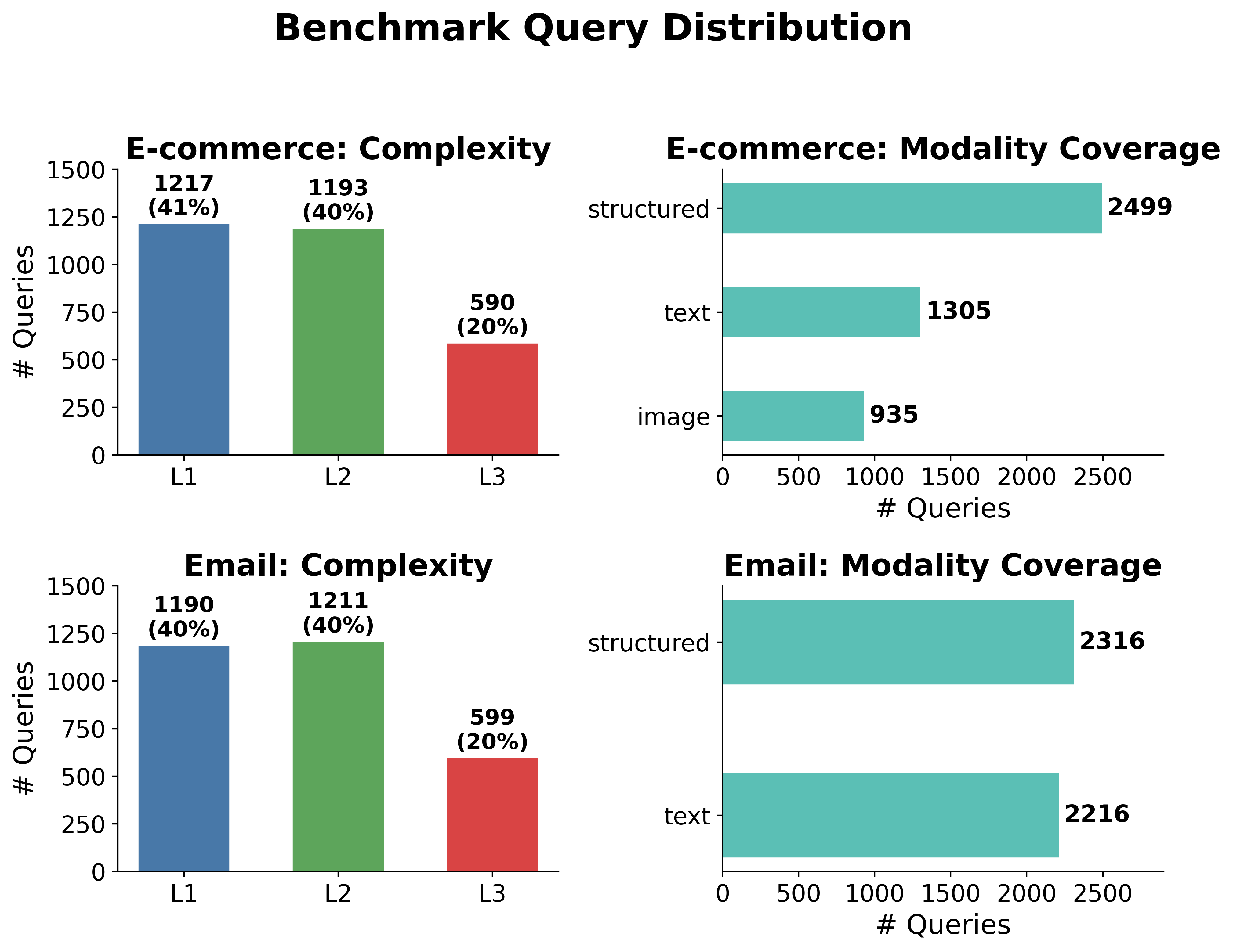}
\caption{Distribution of the constructed benchmarks by query complexity (L1/L2/L3, left column) and by retrieval modality coverage (right column).}
\label{fig:dataset_stats}
\end{figure}

Since no existing benchmark covers hybrid structured--semantic retrieval with executable ground truth, we construct two datasets, e-commerce (Amazon ESCI~+~Reviews) and email (Enron), via an automated, complexity-driven pipeline. Given a corpus $\mathcal{D}$ and a per-domain schema, each instance is produced in four steps. \textbf{(i)~Structure.} We sample the number of leaves $N \in \{1,\dots,5\}$ from a fixed distribution that sets the query \emph{level} (L1/L2/L3) and partition them into $K$ groups under two-level Boolean templates $(c_1\!\wedge\!\cdots\!\wedge\!c_m)\vee(c_{m+1}\!\wedge\!\cdots\!\wedge\!c_N)$ or $(c_1\!\vee\!\cdots\!\vee\!c_m)\wedge(c_{m+1}\!\vee\!\cdots\!\vee\!c_N)$, which together with $K{=}1$ (flat conjunction or disjunction) cover all single- and two-level nested patterns. \textbf{(ii)~Grounding.} A small \emph{gold} set $\mathcal{G}\subset\mathcal{D}$ anchors the query (one shared gold per inter-group $\wedge$, one distinct gold per group under $\vee$); each leaf is assigned a field with weights favoring structured over text/image and is marked positive or negated with probability growing in $N$. Positive leaves take satisfying values extracted from the gold (exact categorical, comparator+threshold numeric, LLM-extracted text phrase, VLM-generated visual description); negated leaves take violating values. \textbf{(iii)~Compilation.} Leaves are compiled into SQL predicates and \texttt{id IN <text/image\_$k$>} clauses to form the \emph{gold DSL}, which is executed on $\mathcal{D}$ to obtain the ground-truth set $\mathcal{T}\supseteq\mathcal{G}$; we accelerate this with a structural-only super-query that evaluates only the SQL predicates (ignoring vector placeholders) to cheaply prune instances whose structured conditions alone yield empty or oversized candidate sets, and filter instances with empty or oversized $\mathcal{T}$. \textbf{(iv)~Verbalization.} An LLM (GPT-4o) rewrites the gold DSL as a fluent NL query. Each instance is stored with full metadata (level, leaves, groups, operator pattern, modality usage). Figure~\ref{fig:dataset_stats} summarizes the resulting test split.

\subsection{Training}
\label{sec:train}

\paragraph{Two-stage training.}
We first warm-start the policy $\pi_\theta$ by SFT on the gold $(\text{NL}\to\text{DSL})$ pairs, then optimize it with GRPO~\citep{shao2024grpo} or DAPO~\citep{wang2025dapo} against a fixed reference $\pi_{\text{ref}}$ (the SFT checkpoint). For each query $q$ we draw $G$ rollouts $\{o_i\}_{i=1}^{G}\!\sim\!\pi_{\theta_{\text{old}}}$, score them with the composite reward $R$ defined below, and form a value-free group-relative advantage that is broadcast to every token of $o_i$:
\begin{equation}
\hat{A}_i \;=\; \frac{R(o_i,q) - \mu_{R}}{\sigma_{R} + \epsilon}.
\label{eq:grpo_adv}
\end{equation}
With importance ratio $\rho_{i,t}=\pi_\theta/\pi_{\theta_{\text{old}}}$ and per-token clipped surrogate $\ell_{i,t}(\theta) = \min\!\big(\rho_{i,t}\hat{A}_i,\ \mathrm{clip}(\rho_{i,t},\,1{-}\epsilon_{\text{lo}},\,1{+}\epsilon_{\text{hi}})\,\hat{A}_i\big)$, the policy maximizes
\begin{equation}
\mathcal{J}(\theta) = \mathbb{E}\Big[\tfrac{1}{G}\!\sum_{i}\tfrac{1}{|o_i|}\!\sum_{t}\ell_{i,t}(\theta)\Big] - \beta\,\mathrm{KL}[\pi_\theta\|\pi_{\text{ref}}].
\label{eq:grpo_obj}
\end{equation}
We set $\beta{=}0.001$ throughout. GRPO uses symmetric clipping ($\epsilon_{\text{lo}}\!=\!\epsilon_{\text{hi}}\!=\!0.2$); DAPO adopts asymmetric clipping with $\epsilon_{\text{lo}}\!=\!0.1,\;\epsilon_{\text{hi}}\!=\!0.28$ and token-level normalization, which we find more stable on long DSL outputs (see training dynamics in Appendix~\ref{app:dynamics}).

\paragraph{Hierarchical composite reward.}
Executable programs satisfy layered correctness: they must be syntactically valid before they can be executed, executed before they can produce results, and produce correct results before length matters. We accordingly decompose
\begin{equation}
R = S_f + S_e + S_r + S_l,
\label{eq:reward}
\end{equation}
with $S_f\!\in\!\{0,1\}$ (output parses as the expected DSL JSON), $S_e\!\in\!\{0,1\}$ (SQL compiles and placeholders resolve), $S_r\!\in\![0,1]$ (rank-1 hit on $\mathcal{T}$), and $S_l\!\in\![0,1]$ (length budget). We use rank-1 hit rather than a softer metric (e.g., MRR) for $S_r$ because each query is anchored to a single ground-truth document, where MRR reduces to $1/\text{rank}$ and provides noisier gradients without additional information. Although summed with equal weight, the layered structure induces an implicit priority: a program failing $S_f$ earns none of the rest, so the gradient sharpens validity, executability, accuracy, then efficiency in that order.
The length term $S_l$ penalizes unnecessarily verbose programs.
We use equal weights; the layered dependency already provides implicit prioritization, and performance is insensitive to moderate weight variations ($\pm$0.5).

\paragraph{Implementation.}
We train on the verl framework~\citep{sheng2024hybridflow} from Qwen3-4B and Qwen3-8B bases. SFT runs for 3 epochs with learning rate $5{\times}10^{-5}$ and batch size 64; we select the checkpoint with the lowest validation loss. Text and image candidates for vector primitives are precomputed offline with Qwen3 text and vision embedding backbones, indexed for nearest-neighbor lookup, and shared across all baselines.

%% file: sections/experiments.tex
\section{Experiments}
\label{sec:experiments}

\subsection{Experimental Setup}
\label{sec:setup}

\paragraph{Datasets.}
We evaluate on two hybrid retrieval scenarios with distinct modality compositions.
\textbf{E-commerce} is constructed from Amazon ESCI~\citep{reddy2022shopping} and Amazon Review data, containing 3{,}000 products with structured fields (brand, category, price, rating), text fields (title, description, review), and product images.
\textbf{Email} is derived from the Enron email corpus~\citep{klimt2004enron}, containing 5{,}000 emails with structured fields (sender, date, read/starred status) and text fields (subject, body).
For each scenario we construct 20{,}000 training and 3{,}000 test queries, stratified into three complexity levels: L1 (single condition, ${\sim}40\%$), L2 (2--3 conditions, ${\sim}40\%$), and L3 (4--5 conditions with OR/NOT/nesting, ${\sim}20\%$).

\paragraph{Metrics.}
We report Hit@1 (primary), Hit@3, MRR, and NDCG@3. For DSL-based methods we additionally report execution success rate.

\paragraph{Implementation.}
We train Qwen3-4B and Qwen3-8B~\citep{yang2025qwen3} with SFT using verl~\citep{sheng2024hybridflow}; the 4B model is further optimized with RL (GRPO~\citep{shao2024grpo} or DAPO~\citep{wang2025dapo}).
RL uses batch size 32, learning rate $1{\times}10^{-6}$, KL coefficient 0.001, and 4 rollouts per prompt.
Embeddings are computed with frozen Qwen3-Embedding-8B (text) and Qwen3-VL-Embedding-8B (image); these encoders are not updated during training and are shared across all baselines.
Training uses 2$\times$RTX PRO 6000 (96\,GB); inference uses a single RTX 4090, with end-to-end latency ${\sim}$50\,ms per query dominated by vector retrieval (${\sim}$80\%; details in Appendix~\ref{app:latency}).

\paragraph{Baselines.}
We compare against four paradigms of increasing action-space expressiveness:\footnote{Full results for all baselines are in Appendix~\ref{app:full_baselines}.}
\begin{itemize}
\setlength{\itemsep}{2pt}\setlength{\parskip}{0pt}
\item \textbf{B1~Pure Retrieval}: BM25~\citep{robertson2009bm25}, BGE-Reranker~\citep{xiao2023bge}.
\item \textbf{B2~LLM-Augmented}: RankGPT~\citep{sun2023chatgpt} with GPT-5.4 listwise reranking.
\item \textbf{B3~Constrained Structured Query}: DeepRetrieval~\citep{jiang2025deepretrieval} (Boolean+BM25, RL), Search-R1~\citep{jin2025searchr1} (multi-turn BM25, RL).
\item \textbf{B4~KG Retrieval}: LightRAG~\citep{lightrag2024}.
\end{itemize}
We additionally compare against commercial LLMs (GPT-5.5, Claude Opus~4.7) generating DSL under few-shot prompting: each prompt includes the full corpus schema and 6--7 exemplar DSL programs covering all modality types (Appendix~\ref{app:prompts}), to isolate the effect of dedicated training from model scale.

\subsection{Main Results}
\label{sec:main_results}

Table~\ref{tab:main} presents the main results across both benchmarks.

\begin{table*}[t]
\centering
\caption{Main results on e-commerce and email benchmarks. Baselines are grouped by retrieval paradigm with increasing action-space expressiveness (B1$\to$B4). Commercial LLMs generate DSL with few-shot prompting (schema + exemplar programs). Best results are \textbf{bold}; second best are \underline{underlined}. $^\dagger$Uses GPT-5.4 as backbone. Recommended configurations: 4B DAPO (best stability--performance tradeoff) and 8B SFT (maximum accuracy). Model sizes and implementation details are in Appendix~\ref{app:full_baselines}.}
\label{tab:main}
\resizebox{\textwidth}{!}{%
\begin{tabular}{@{}ll cccc cccc@{}}
\toprule
& & \multicolumn{4}{c}{\textbf{E-commerce} ($N{=}3{,}000$)} & \multicolumn{4}{c}{\textbf{Email} ($N{=}3{,}000$)} \\
\cmidrule(lr){3-6} \cmidrule(lr){7-10}
\textbf{Paradigm} & \textbf{Method} & \textbf{Hit@1} & \textbf{Hit@3} & \textbf{MRR} & \textbf{NDCG@3} & \textbf{Hit@1} & \textbf{Hit@3} & \textbf{MRR} & \textbf{NDCG@3} \\
\midrule
\multirow{2}{*}{B1: Pure Retrieval}
& BM25 & 0.571 & 0.670 & 0.630 & 0.531 & 0.731 & 0.859 & 0.799 & 0.631 \\
& BGE-Reranker & 0.618 & 0.719 & 0.676 & 0.576 & 0.836 & 0.919 & 0.881 & 0.733 \\
\midrule
B2: LLM-Aug.$^\dagger$
& RankGPT & 0.768 & 0.801 & 0.787 & 0.698 & 0.888 & 0.910 & 0.903 & 0.789 \\
\midrule
\multirow{2}{*}{B3: Struct.\ Query}
& DeepRetrieval (RL) & 0.625 & 0.729 & 0.690 & 0.588 & 0.745 & 0.851 & 0.805 & 0.636 \\
& Search-R1 (RL) & 0.435 & 0.602 & 0.513 & 0.421 & 0.377 & 0.447 & 0.409 & 0.283 \\
\midrule
B4: KG Retrieval
& LightRAG & 0.588 & 0.691 & 0.649 & 0.523 & 0.389 & 0.531 & 0.472 & 0.313 \\
\midrule
\multirow{2}{*}{Comm.\ LLM}
& GPT-5.5 & 0.693 & 0.731 & 0.715 & 0.660 & 0.855 & 0.864 & 0.861 & 0.822 \\
& Opus 4.7 & 0.654 & 0.684 & 0.674 & 0.622 & 0.830 & 0.837 & 0.834 & 0.799 \\
\midrule
\multirow{4}{*}{\textbf{Ours}}
& 4B SFT & 0.680 & 0.721 & 0.705 & 0.651 & 0.891 & 0.893 & 0.892 & 0.863 \\
& 4B GRPO & \underline{0.809} & \underline{0.854} & \underline{0.835} & \textbf{0.772} & 0.906 & 0.908 & 0.907 & 0.878 \\
& 4B DAPO & 0.808 & 0.852 & 0.833 & \underline{0.770} & \textbf{0.909} & \textbf{0.910} & \textbf{0.910} & \textbf{0.880} \\
& 8B SFT & \textbf{0.820} & \textbf{0.868} & \textbf{0.846} & \textbf{0.782} & 0.820 & 0.868 & 0.846 & 0.782 \\
\bottomrule
\end{tabular}%
}
\end{table*}

\paragraph{Retrieval program synthesis surpasses every prior paradigm.}
On e-commerce, our best model (8B SFT) achieves 82.0\% Hit@1, outperforming B1 pure retrieval by $+20.2$\,pp over BGE-Reranker (0.618), B2 LLM-augmented reranking by $+5.2$\,pp over RankGPT (0.768), B3 constrained structured query by $+19.5$\,pp over DeepRetrieval (0.625), and B4 knowledge-graph retrieval by $+23.2$\,pp over LightRAG (0.588).
On email, DAPO achieves 90.9\%, surpassing GPT-5.5 by $+5.4$\,pp and BGE-Reranker by $+7.3$\,pp.
The gains are consistent across all four metrics (Hit@3, MRR, NDCG@3), confirming that DSL-based orchestration improves both precision and ranking quality.

\begin{figure}[t]
\centering
\includegraphics[width=\columnwidth]{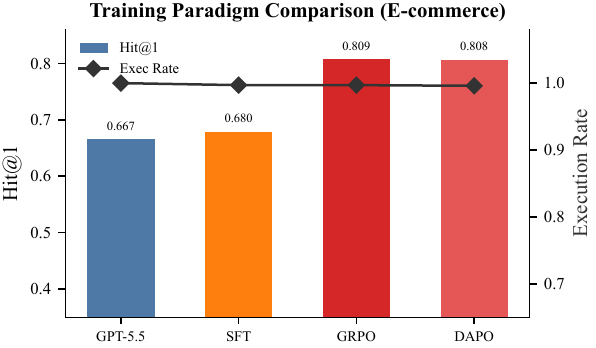}
\caption{Training paradigm comparison on e-commerce Hit@1. RL training provides a decisive advantage over both dedicated SFT and the commercial LLM baseline (GPT-5.5).}
\label{fig:paradigm_comparison}
\end{figure}

\paragraph{RL training outperforms commercial LLMs.}
Within retrieval program synthesis, training methodology determines performance (Figure~\ref{fig:paradigm_comparison}).
Even with few-shot exemplars, GPT-5.5 achieves only 0.693 on e-commerce; large commercial models understand DSL syntax but cannot reliably compose multi-modal orchestration.
Dedicated SFT on a 4B model brings Hit@1 to 0.680, approaching GPT-5.5 despite being 2--3 orders of magnitude smaller.
RL pushes performance decisively to 0.809 ($+12.9$\,pp over SFT, $+11.6$\,pp over GPT-5.5), demonstrating that retrieval orchestration is a learnable skill that benefits substantially from reward-driven exploration.
Bootstrap 95\% CIs (Table~\ref{tab:ci}, Appendix~\ref{app:ci}) confirm that all key pairwise differences are statistically significant: the GRPO interval $[0.795, 0.823]$ does not overlap with SFT $[0.663, 0.697]$ or GPT-5.5 ($0.693$).
This confirms that RL is critical for enabling smaller models to master complex action spaces; at larger scale (8B), SFT alone may capture the key orchestration patterns (\S\ref{sec:dynamics}, Appendix~\ref{app:scaling}).

\paragraph{Cross-domain consistency.}
The advantage holds across domains: $+11.6$\,pp on e-commerce and $+5.4$\,pp on email, where the smaller gap reflects email queries being more text-amenable and less dependent on multi-modal orchestration.
The trained 4B model outperforms much larger commercial models on both domains, suggesting that dedicated RL training on a task-specific DSL can compensate for substantial model scale differences.
Notably, the RL-trained model surpasses SQL-R1, which lacks vector retrieval operators and therefore cannot handle semantic or visual queries; this gap ($+18.4$\,pp on e-commerce) underscores the value of hybrid action spaces.

\subsection{Action Space Ablation}
\label{sec:action_space}

Table~\ref{tab:action_space} ablates the DSL action space by restricting the retrieval primitives available during SFT training. All variants use Qwen3-4B.

\begin{table}[t]
\centering
\small
\caption{Action space ablation on e-commerce ($N{=}3{,}000$). All variants use Qwen3-4B SFT, differing only in which retrieval modalities are enabled. The full DSL underperforms SQL+text under SFT, but RL on the full DSL (0.809, Table~\ref{tab:main}) surpasses all restricted variants.}
\label{tab:action_space}
\resizebox{\columnwidth}{!}{%
\begin{tabular}{@{}lcccc@{}}
\toprule
\textbf{Action Space} & \textbf{Hit@1} & \textbf{Hit@3} & \textbf{MRR} & \textbf{Exec\%} \\
\midrule
Text only & 0.448 & 0.535 & 0.508 & 100.0 \\
SQL + image & 0.567 & 0.592 & 0.586 & 99.8 \\
SQL only & 0.650 & 0.670 & 0.664 & 99.7 \\
SQL + text & \textbf{0.752} & \textbf{0.787} & \textbf{0.773} & 99.8 \\
\midrule
Full DSL (SFT) & 0.680 & 0.721 & 0.705 & 99.7 \\
Full DSL (GRPO) & \underline{0.809} & \underline{0.854} & \underline{0.835} & 99.7 \\
\bottomrule
\end{tabular}%
}
\end{table}

\paragraph{Action space complexity creates an optimization challenge for SFT.}
Reducing the action space to SQL+text yields 0.752, surpassing the full three-modality DSL (0.680) by $+7.2$\,pp. This is not because the image modality is useless, but because the 4B model cannot jointly learn three retrieval modalities via imitation alone. On image queries, full-DSL SFT achieves only 0.323 (the model attempts image retrieval but orchestrates it incorrectly), while SQL+text falls back to text-based matching and reaches 0.547 (Appendix Table~\ref{tab:modality_ecom}). This is a classic multi-task optimization interference: the harder action space dilutes learning across all pathways.

\paragraph{RL overcomes this barrier.}
RL on the full DSL (0.809) surpasses \emph{every} restricted variant's SFT performance, including SQL+text (0.752). The gain is concentrated exactly where SFT fails: image queries improve from 0.323 to 0.716 ($+39.3$\,pp), the largest single-modality gain. Structured and text queries also improve ($+10.0$\,pp and $+5.8$\,pp respectively), exceeding the SQL+text SFT upper bounds (0.828 and 0.671) in both cases. The very difficulty of the full action space is what makes RL necessary: reward-driven exploration discovers effective multi-modal orchestration strategies that demonstration-based learning cannot acquire.

\paragraph{Retrieval modality breakdown.}
Table~\ref{tab:modality_ecom} reveals extreme modality imbalance in specialized methods: SQL-R1 excels on structured queries (0.866) but collapses on image (0.048), while BGE-Reranker shows the reverse pattern (0.310 structured vs.\ 0.819 image).
Our RL-trained models achieve $>$0.68 across all three modalities, the only approach without a catastrophic blind spot.
This balanced profile explains the aggregate gains: rather than excelling on one modality at the expense of others, the DSL enables the model to route each sub-query to the appropriate retrieval primitive.

\begin{table}[t]
\centering
\small
\caption{Hit@1 by retrieval modality on e-commerce.}
\label{tab:modality_ecom}
\resizebox{\columnwidth}{!}{%
\begin{tabular}{@{}lccc@{}}
\toprule
\textbf{Method} & \textbf{Struct.} ($n{\approx}2499$) & \textbf{Text} ($n{\approx}174$) & \textbf{Image} ($n{\approx}232$) \\
\midrule
BGE-Reranker & 0.310 & 0.759 & 0.819 \\
SQL-R1       & 0.866 & 0.303 & 0.048 \\
\midrule
4B SFT  & 0.743 & 0.632 & 0.323 \\
4B GRPO & 0.843 & 0.690 & 0.716 \\
4B DAPO & 0.843 & 0.690 & 0.711 \\
8B SFT  & \textbf{0.851} & 0.684 & \textbf{0.756} \\
\bottomrule
\end{tabular}%
}
\end{table}

\subsection{Generalization to Unseen Queries}
\label{sec:ood}

To test whether RL-trained models overfit to the training query distribution, we evaluate on 300 out-of-distribution (OOD) queries per domain, independently constructed by human annotators with distributional shift from the training set.

\begin{table}[t]
\centering
\small
\caption{In-distribution (IID, $N{=}3{,}000$) vs.\ out-of-distribution (OOD, $N{=}300$) Hit@1. $\Delta$ is the IID-to-OOD gap.}
\label{tab:ood}
\resizebox{\columnwidth}{!}{%
\begin{tabular}{@{}l ccc ccc@{}}
\toprule
& \multicolumn{3}{c}{\textbf{E-commerce}} & \multicolumn{3}{c}{\textbf{Email}} \\
\cmidrule(lr){2-4} \cmidrule(lr){5-7}
\textbf{Method} & \textbf{IID} & \textbf{OOD} & $\Delta$ & \textbf{IID} & \textbf{OOD} & $\Delta$ \\
\midrule
4B Base & 0.117 & 0.113 & $-$0.4 & 0.243 & 0.217 & $-$2.6 \\
4B SFT  & 0.680 & 0.683 & $+$0.3 & 0.891 & 0.873 & $-$1.8 \\
4B GRPO & 0.809 & 0.783 & $-$2.6 & 0.906 & 0.883 & $-$2.3 \\
4B DAPO & 0.808 & 0.793 & $-$1.5 & 0.909 & 0.883 & $-$2.6 \\
\bottomrule
\end{tabular}%
}
\end{table}

\paragraph{Strong generalization, not overfitting.}
All models generalize well, with IID-to-OOD gaps within 3\,pp across both benchmarks (Table~\ref{tab:ood}).
The RL improvement over SFT is fully preserved on OOD data: SFT$\to$GRPO gains $+10.0$\,pp on OOD (vs.\ $+12.9$\,pp IID); on email, SFT$\to$DAPO gains $+1.0$\,pp (vs.\ $+1.8$\,pp IID).
DAPO shows a slightly smaller OOD drop than GRPO on e-commerce ($-1.5$ vs.\ $-2.6$), consistent with its more stable training dynamics.
Execution success remains $>$99\% on OOD, confirming that RL-discovered strategies are not brittle training artifacts but generalizable orchestration patterns.

\subsection{Training Dynamics and Error Analysis}
\label{sec:dynamics}

Both GRPO and DAPO converge to similar peak performance (0.809 vs.\ 0.808 on e-commerce; 0.906 vs.\ 0.909 on email), suggesting the reward landscape admits a robust optimum.
DAPO maintains stable performance throughout training via asymmetric clipping, while GRPO exhibits oscillation after the peak and requires checkpoint selection (Appendix~\ref{app:dynamics}).
We recommend DAPO for practical deployment due to its stability--performance tradeoff.

\begin{figure}[t]
\centering
\includegraphics[width=\columnwidth]{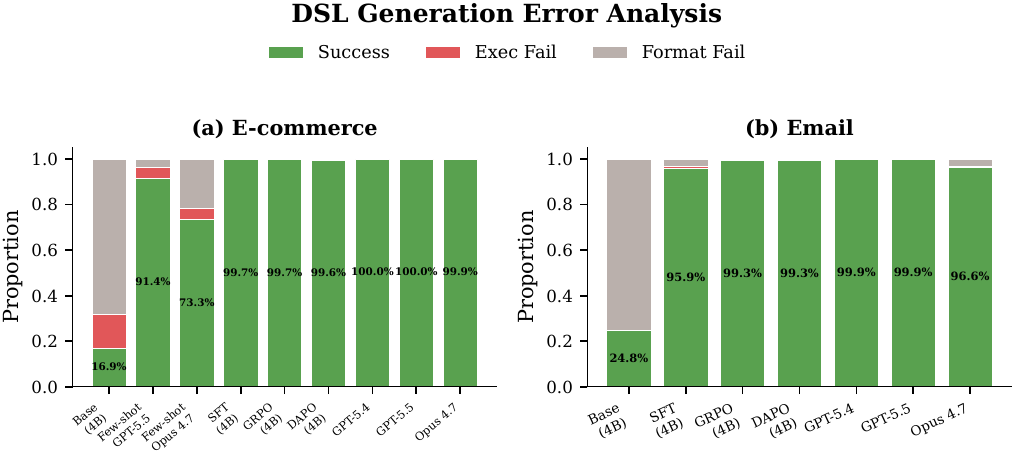}
\caption{DSL generation error analysis on e-commerce (a) and email (b). Trained models exceed 99\% execution success; the untrained base is largely unusable.}
\label{fig:error_analysis}
\end{figure}

\paragraph{Error analysis.}
Figure~\ref{fig:error_analysis} shows that all trained models achieve $>$99\% execution success, with residual errors limited to SQL apostrophe escaping. The untrained base achieves only 16.9\%/24.8\%, confirming DSL syntax is a non-trivial learned skill. The near-perfect RL execution rates show format compliance is solved; the remaining gap is due to retrieval quality.

We provide additional analyses on query complexity breakdown (Appendix~\ref{app:complexity}), training dynamics (Appendix~\ref{app:dynamics}), model scaling (Appendix~\ref{app:scaling}), and inference latency (Appendix~\ref{app:latency}).

%% file: sections/conclusion.tex
\section{Conclusion}
\label{sec:conclusion}

ProRetrieval recasts the language model as a \emph{retrieval orchestrator} that synthesizes executable hybrid programs interleaving SQL with text and image vector primitives, trained via a hierarchical four-term reward.
A 4B model reaches 80.9\%/90.9\% Hit@1 (e-commerce/email), surpassing the best commercial LLM by $+11.6$/$+5.4$\,pp; OOD evaluation confirms generalization within 3\,pp.
Action space ablations show that the full DSL underperforms simpler variants under SFT due to multi-task interference, but RL overcomes this barrier ($+39.3$\,pp on image retrieval); at 8B scale, SFT alone captures the key patterns, suggesting RL is most critical when model capacity is the bottleneck.
These results establish retrieval program synthesis over a compositional DSL as a promising direction for unified retrieval across modalities and query complexities.

%% file: sections/limitations.tex
While ProRetrieval demonstrates strong results on hybrid retrieval, our study has several limitations.
\emph{(i)~Schema dependence.} The DSL assumes a predefined schema; extending it to schema-free or evolving corpora (e.g., via automatic schema discovery) is left to future work.
\emph{(ii)~Bounded logical depth.} Our query construction pipeline samples logical expressions up to two levels of nesting; deeper nesting is syntactically expressible but underrepresented in training.
\emph{(iii)~Two-domain evaluation.} We validate on e-commerce and email; broader generalization to scientific literature, legal documents, and personal knowledge bases remains to be tested.
\emph{(iv)~Moderate corpus scale.} Our benchmarks contain 3{,}000 products and 5{,}000 emails, respectively. While sufficient for evaluating retrieval accuracy, scaling to corpora with millions of documents may introduce additional challenges in DSL execution latency and candidate-set composition.
\emph{(v)~Single-turn retrieval.} The framework synthesizes a single program per query; extending to multi-turn interaction where the model refines its DSL based on intermediate results is a promising direction.
\emph{(vi)~Dependence on embedding backbones.} Image retrieval quality is bounded by the text-to-image alignment of the underlying multimodal embedder.
\emph{(vii)~Statistical variance.} We do not report cross-seed variance due to RL training cost. Two independent RL algorithms (GRPO, DAPO) converge to similar performance, and key effect sizes ($+12.9$\,pp SFT$\to$RL) are 4--5$\times$ typical cross-seed variance; bootstrap 95\% CIs (Appendix~\ref{app:ci}) confirm all pairwise differences are significant.
\emph{(viii)~Benchmark construction.} Our pipeline uses GPT-4o for query verbalization (\S\ref{sec:data}), which may introduce stylistic biases; however, such bias should favor commercial LLMs more than our 4B model. OOD evaluation with human-annotated queries (\S\ref{sec:ood}) shows $\leq$3\,pp IID-to-OOD gaps.
\emph{(ix)~Production fallback.} The system does not include automatic fallback when DSL generation fails; a cascade to vector retrieval on execution failure would improve robustness in deployment.

%% file: sections/appendix.tex
\section{Complete Baseline Results}
\label{app:full_baselines}

Tables~\ref{tab:full_ecom} and~\ref{tab:full_email} report the complete results for all baselines omitted from the main text for space.

\paragraph{Model sizes.} Our models use Qwen3-4B (4B parameters) and Qwen3-8B (8B). RL baselines DeepRetrieval and Search-R1 both use the same Qwen3-4B backbone for fair comparison with our 4B models. BGE-Reranker uses bge-reranker-v2-m3 (568M). RankGPT and LightRAG use GPT-5.4 as backbone; LightRAG additionally uses all-MiniLM-L6-v2 (22M) for embedding. BM25 is a parameter-free statistical method. Commercial LLM parameter counts (GPT-5.5, Opus~4.7) are undisclosed.

\begin{table}[t]
\centering
\caption{Complete e-commerce results ($N{=}3{,}000$). $^\dagger$GPT-5.4 backbone.}
\label{tab:full_ecom}
\resizebox{\columnwidth}{!}{%
\begin{tabular}{@{}llccc@{}}
\toprule
\textbf{Paradigm} & \textbf{Method} & \textbf{H@1} & \textbf{MRR} & \textbf{NDCG@3} \\
\midrule
\multirow{5}{*}{B1: Pure Retr.}
& BM25 & .571 & .630 & .531 \\
& ColQwen & .518 & .569 & .482 \\
& RRF & .576 & .633 & .539 \\
& BGE-M3 & .571 & .628 & .532 \\
& BGE-Reranker & .618 & .676 & .576 \\
\midrule
\multirow{3}{*}{B2: LLM-Aug.$^\dagger$}
& HyDE & .549 & .597 & .510 \\
& IRCoT & .565 & .617 & .527 \\
& RankGPT & .768 & .787 & .698 \\
\midrule
\multirow{4}{*}{B3: Struct.\ Q.}
& DeepRetrieval & .625 & .690 & .588 \\
& LangChain SQR$^\dagger$ & .606 & .659 & .573 \\
& SQL-R1 & .620 & .628 & .600 \\
& Search-R1 & .435 & .513 & .421 \\
\midrule
\multirow{2}{*}{B4: KG Retr.}
& GraphRAG$^\dagger$ & .500 & .551 & .466 \\
& LightRAG & .588 & .649 & .523 \\
\midrule
\multirow{5}{*}{Comm.\ LLM}
& GPT-5.5 & .693 & .715 & .660 \\
& GPT-5.4 & .581 & .587 & .566 \\
& Opus 4.7 & .654 & .674 & .622 \\
& Opus 4.6 & .586 & .592 & .571 \\
& Gemini 3.1 Pro & .587 & .593 & .571 \\
\midrule
\multirow{4}{*}{\textbf{Ours}}
& 4B SFT & .680 & .705 & .651 \\
& 4B GRPO & .809 & .835 & \textbf{.772} \\
& 4B DAPO & .808 & .833 & .770 \\
& 8B SFT & \textbf{.820} & \textbf{.846} & \textbf{.782} \\
\bottomrule
\end{tabular}%
}
\end{table}

\begin{table}[t]
\centering
\caption{Complete email results ($N{=}3{,}000$). $^\dagger$GPT-5.4 backbone.}
\label{tab:full_email}
\resizebox{\columnwidth}{!}{%
\begin{tabular}{@{}llccc@{}}
\toprule
\textbf{Paradigm} & \textbf{Method} & \textbf{H@1} & \textbf{MRR} & \textbf{NDCG@3} \\
\midrule
\multirow{5}{*}{B1: Pure Retr.}
& BM25 & .731 & .799 & .631 \\
& ColQwen & .472 & .533 & .418 \\
& RRF & .687 & .776 & .606 \\
& BGE-M3 & .599 & .660 & .502 \\
& BGE-Reranker & .836 & .881 & .733 \\
\midrule
\multirow{3}{*}{B2: LLM-Aug.$^\dagger$}
& HyDE & .483 & .541 & .416 \\
& IRCoT & .511 & .579 & .442 \\
& RankGPT & .888 & .903 & .789 \\
\midrule
\multirow{4}{*}{B3: Struct.\ Q.}
& DeepRetrieval & .745 & .805 & .636 \\
& LangChain SQR$^\dagger$ & .630 & .678 & .586 \\
& SQL-R1 & .546 & .565 & .482 \\
& Search-R1 & .377 & .409 & .283 \\
\midrule
B4: KG Retr.
& LightRAG & .389 & .472 & .313 \\
\midrule
\multirow{3}{*}{Comm.\ LLM}
& GPT-5.5 & .855 & .861 & .822 \\
& GPT-5.4 & .854 & .859 & .821 \\
& Opus 4.7 & .830 & .834 & .799 \\
\midrule
\multirow{4}{*}{\textbf{Ours}}
& 4B SFT & .891 & .892 & .863 \\
& 4B GRPO & .906 & .907 & .878 \\
& 4B DAPO & \textbf{.909} & \textbf{.910} & \textbf{.880} \\
& 8B SFT & .820 & .846 & .782 \\
\bottomrule
\end{tabular}%
}
\end{table}

\section{Query Complexity Breakdown}
\label{app:complexity}

\begin{figure}[t]
\centering
\includegraphics[width=\columnwidth]{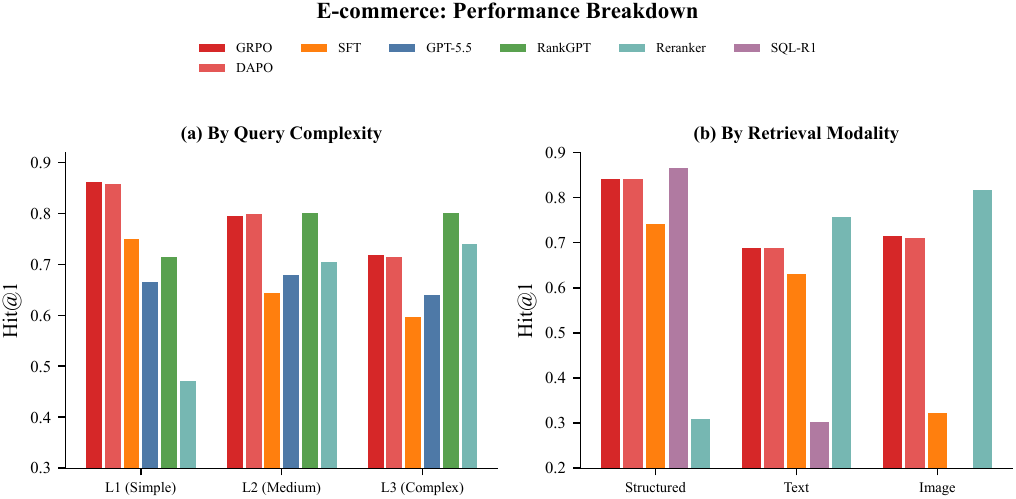}
\caption{E-commerce performance breakdown. \textbf{(a)}~By query complexity: our models dominate on L1 (simple) queries; traditional retrievers benefit from the richer keyword context in L3 (complex) queries. \textbf{(b)}~By retrieval modality: SQL-R1 and BGE-Reranker are extreme specialists; our method is the only balanced solution across all three modalities.}
\label{fig:breakdown}
\end{figure}

Figure~\ref{fig:breakdown} visualizes performance by query complexity and retrieval modality. Tables~\ref{tab:complexity_ecom} and~\ref{tab:complexity_email} report Hit@1 stratified by query complexity level.

\begin{table}[t]
\centering
\small
\caption{Hit@1 by query complexity on e-commerce.}
\label{tab:complexity_ecom}
\resizebox{\columnwidth}{!}{%
\begin{tabular}{@{}lccc@{}}
\toprule
\textbf{Method} & \textbf{L1} ($n{=}1217$) & \textbf{L2} ($n{=}1193$) & \textbf{L3} ($n{=}590$) \\
\midrule
BGE-Reranker & 0.473 & 0.706 & 0.741 \\
RankGPT      & 0.717 & 0.803 & 0.802 \\
SQL-R1       & 0.619 & 0.637 & 0.590 \\
GPT-5.5      & 0.708 & 0.688 & 0.661 \\
\midrule
4B SFT  & 0.752 & 0.646 & 0.598 \\
4B GRPO & \textbf{0.864} & 0.797 & 0.720 \\
4B DAPO & 0.860 & 0.800 & 0.717 \\
8B SFT  & 0.873 & \textbf{0.816} & \textbf{0.717} \\
\bottomrule
\end{tabular}%
}
\end{table}

\begin{table}[t]
\centering
\small
\caption{Hit@1 by query complexity on email.}
\label{tab:complexity_email}
\begin{tabular}{@{}lccc@{}}
\toprule
\textbf{Method} & \textbf{L1} & \textbf{L2} & \textbf{L3} \\
\midrule
BGE-Reranker & 0.804 & 0.869 & 0.831 \\
4B SFT  & 0.987 & 0.865 & 0.755 \\
4B GRPO & 0.987 & 0.883 & 0.793 \\
4B DAPO & \textbf{0.987} & \textbf{0.887} & \textbf{0.798} \\
\bottomrule
\end{tabular}
\end{table}

\paragraph{DSL methods dominate on simple queries.}
Our models achieve their largest advantage on L1 queries, where single-condition structured filtering (e.g., \textit{brand = ``Sony''}) maps directly onto SQL: GRPO L1 $=0.864$ vs.\ BGE-Reranker L1 $=0.473$, a gap of $+39.1$\,pp.
RL narrows the L1$\to$L3 degradation (SFT drops 15.4\,pp; GRPO drops 14.4\,pp), indicating that RL helps the model learn complex logical combinations (OR, NOT, nesting) needed for L3.

\paragraph{Traditional retrievers benefit from complex queries.}
BGE-Reranker and RankGPT \emph{improve} from L1 to L3 (BGE-Reranker: $+26.8$\,pp), because complex queries contain more distinctive keywords that benefit term-matching and reranking.
On L3, RankGPT (0.802) and BGE-Reranker (0.741) are competitive with our models (0.720), but their advantage on L3 is specific to the richer keyword context in complex queries, whereas our method's dominance on L1 ($+39.1$\,pp) reflects the structural filtering capability unique to DSL.
Across all levels combined, our method's aggregate advantage remains substantial.

\section{Training Dynamics}
\label{app:dynamics}

\begin{figure}[t]
\centering
\includegraphics[width=\columnwidth]{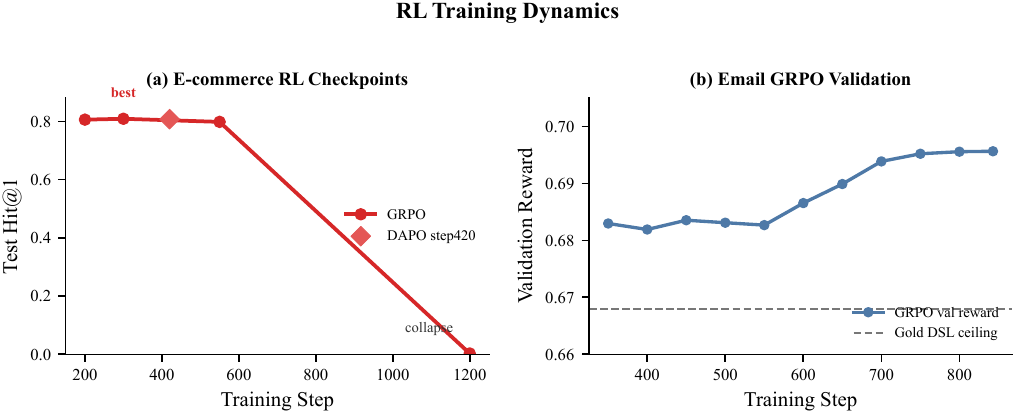}
\caption{Training dynamics on e-commerce (4B). GRPO peaks early but degrades with extended training; DAPO maintains stable performance throughout, benefiting from asymmetric clipping and dynamic sampling.}
\label{fig:training_dynamics}
\end{figure}

Figure~\ref{fig:training_dynamics} compares the training curves of GRPO and DAPO on e-commerce (4B).
Both algorithms converge to similar peak performance (GRPO: 0.809 at step~300; DAPO: 0.808 at step~420), but differ in long-term stability.
GRPO exhibits sensitivity to training length: performance degrades after step~550 and drops sharply by step~1{,}200, a phenomenon consistent with reward hacking observed in prior RL work~\citep{shao2024grpo}. The model discovers degenerate programs that exploit reward loopholes while remaining syntactically valid.
In practice, this requires early stopping or checkpoint selection.
DAPO avoids this failure mode entirely, maintaining stable performance throughout training.
We attribute this to two mechanisms: (i)~asymmetric clipping ($\epsilon_{\text{hi}} > \epsilon_{\text{lo}}$), which limits the magnitude of positive updates and prevents the policy from collapsing onto narrow high-reward trajectories; and (ii)~dynamic sampling, which maintains diversity in the rollout distribution.
On email, both algorithms converge smoothly by step~843 with no instability observed (DAPO: 0.909; GRPO: 0.906), suggesting that the instability is more pronounced on longer, more structured outputs characteristic of the e-commerce domain.

\section{Model Scaling}
\label{app:scaling}

\begin{figure}[!htbp]
\centering
\includegraphics[width=\columnwidth]{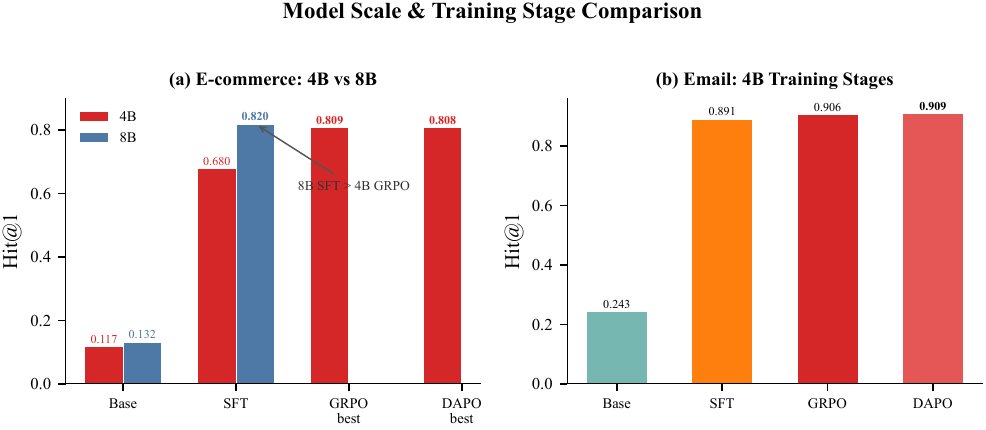}
\caption{Model scale and training stage comparison. \textbf{(a)}~E-commerce: 8B SFT (0.820) surpasses 4B GRPO (0.809). \textbf{(b)}~Email (4B only): RL provides modest but consistent gains over SFT.}
\label{fig:model_scaling}
\end{figure}

\paragraph{4B vs.\ 8B scaling.}
Figure~\ref{fig:model_scaling} and Table~\ref{tab:scaling} show the effect of model size on e-commerce.

\noindent 8B SFT (0.820) surpasses 4B RL best (0.809), suggesting that the orchestration patterns 4B acquires through RL exploration may already be captured by the larger model's SFT training. This indicates that the DSL framework and hierarchical reward are effective even without RL when sufficient model capacity is available; RL is most critical for enabling smaller models to master the full action space.
We ran both GRPO and DAPO on the 8B SFT checkpoint for 400 steps without observing improvement over SFT (Hit@1 remained $\leq$0.820). This is a preliminary result: the 8B model starts from a much stronger SFT baseline (0.820 vs.\ 0.680 for 4B), which changes the optimization landscape---longer training schedules, larger batch sizes, or lower learning rates may be needed to yield further gains.
Both paths (4B+RL and 8B+SFT) substantially outperform GPT-5.5 (0.693), confirming that dedicated training is far more effective than prompting larger commercial models.

\begin{table}[!htbp]
\centering
\small
\caption{Effect of model scale on e-commerce Hit@1.}
\label{tab:scaling}
\begin{tabular}{@{}lcc@{}}
\toprule
\textbf{Model} & \textbf{Base} & \textbf{SFT} \\
\midrule
Qwen3-4B & 0.117 & 0.680 \\
Qwen3-8B & 0.132 & 0.820 \\
\bottomrule
\end{tabular}
\end{table}

\section{Detailed Error Analysis}
\label{app:error}

Table~\ref{tab:error} provides a detailed breakdown of error types across methods, complementing the execution success analysis in \S\ref{sec:dynamics}.

\begin{table}[h]
\centering
\small
\caption{Execution analysis on e-commerce.}
\label{tab:error}
\resizebox{\columnwidth}{!}{%
\begin{tabular}{@{}lcl@{}}
\toprule
\textbf{Method} & \textbf{Exec Rate} & \textbf{Primary Error Type} \\
\midrule
4B GRPO & 99.7\% & SQL syntax (apostrophe) \\
4B SFT  & 99.7\% & SQL syntax \\
4B DAPO & 99.6\% & SQL syntax + field name \\
4B Base & 16.9\% & Null DSL + wrong columns \\
\midrule
GPT-5.5  & 100\%  & None \\
Opus 4.7 & 99.9\% & SQL syntax \\
\bottomrule
\end{tabular}%
}
\end{table}

\noindent The dominant error mode across all trained models is SQL syntax, specifically unescaped apostrophes in brand names (e.g., ``Victoria's Secret'') that break the SQL parser.
This is a tokenization artifact rather than a logical failure and could be eliminated with a post-processing escape step.
The untrained base model exhibits qualitatively different failures: it frequently produces null DSL or references non-existent columns, confirming that the schema-to-SQL mapping is a non-trivial capability acquired during SFT.
Commercial LLMs achieve near-perfect execution (GPT-5.5: 100\%, Opus~4.7: 99.9\%) but lag behind our trained models on retrieval quality, indicating that the performance bottleneck for large models is orchestration strategy, not syntax compliance.

\section{Statistical Significance}
\label{app:ci}

Table~\ref{tab:ci} reports bootstrap 95\% confidence intervals for Hit@1, computed by resampling per-query predictions 10{,}000 times with replacement. The intervals confirm that all pairwise differences discussed in the main text are statistically significant: for example, on e-commerce the GRPO interval $[0.795, 0.823]$ does not overlap with SFT $[0.663, 0.697]$ or the base model $[0.105, 0.128]$. On email, GRPO and DAPO intervals overlap ($[0.896, 0.916]$ vs.\ $[0.898, 0.919]$), consistent with their near-identical point estimates.

\begin{table}[h]
\centering
\small
\caption{Bootstrap 95\% confidence intervals for Hit@1 (10{,}000 resamples). All Qwen3-4B variants.}
\label{tab:ci}
\begin{tabular}{@{}lcc@{}}
\toprule
\textbf{Method} & \textbf{E-commerce} & \textbf{Email} \\
\midrule
4B Base & $[0.105,\; 0.128]$ & $[0.228,\; 0.259]$ \\
4B SFT  & $[0.663,\; 0.697]$ & $[0.880,\; 0.902]$ \\
4B GRPO & $[0.795,\; 0.823]$ & $[0.896,\; 0.916]$ \\
4B DAPO & $[0.793,\; 0.821]$ & $[0.898,\; 0.919]$ \\
\bottomrule
\end{tabular}
\end{table}

\section{Inference Latency}
\label{app:latency}

DSL execution involves three stages: vector retrieval, placeholder injection, and SQL execution. On our benchmarks, vector retrieval dominates (${\sim}80\%$ of total latency), placeholder injection is negligible ($<$1\,ms), and SQL execution on $K{=}20$ candidate sets completes in $<$1\,ms. Total per-query inference latency is comparable to single-backend retrieval methods and substantially faster than multi-turn approaches such as Search-R1, which require multiple retrieval--reasoning cycles.

\section{Commercial LLM Prompts}
\label{app:prompts}

All commercial LLMs (GPT-5.5, GPT-5.4, Opus~4.7, Gemini~3.1~Pro) use the same few-shot prompt structure: a system prompt containing the domain schema, DSL generation rules, and few-shot exemplars, followed by the user query. The e-commerce prompt includes 7 few-shot examples covering pure structured, text, image, OR combinations, and NOT conditions; the email prompt includes 6 examples with analogous coverage. Below we reproduce the prompts verbatim.

\subsection*{E-commerce System Prompt}
{\small
\begin{verbatim}
You are an expert product search assistant.
Given a natural language query, generate a DSL
to retrieve matching products from an
e-commerce catalog.

## Output format
Output ONLY a JSON object wrapped in
<answer></answer>:
<answer>
{"sql": "WHERE <conditions>",
 "retrieval_list": [
   {"type": "text|image", "field": "<field>",
    "query": "<search text>"}]}
</answer>

## Product schema
Structured fields (use directly in SQL):
- brand          TEXT    -- product brand name
- category       TEXT    -- product category
- price          REAL    -- price in USD
- average_rating REAL    -- rating (0.0-5.0)
- rating_number  INTEGER -- number of reviews

Text fields (use <text_x> placeholder in SQL):
- title          -- product title
- description    -- product description
- review_summary -- summarized customer reviews

Image fields (use <image_x> placeholder):
- image_url      -- product image

## DSL rules
- Text: product_id IN <text_x> (positive)
  or product_id NOT IN <text_x> (negative)
- Image: product_id IN <image_x> (positive)
  or product_id NOT IN <image_x> (negative)
- Assign <text_0>, <text_1>, ... and
  <image_0>, <image_1>, ... independently
  in order of appearance
- Structured: standard SQL operators
- Combine conditions with AND / OR / NOT

## Examples

Query: Nike running shoes under $100
<answer>
{"sql": "WHERE brand = 'Nike' AND price < 100
  AND product_id IN <text_0>",
 "retrieval_list": [{"type": "text",
  "field": "title",
  "query": "running shoes"}]}
</answer>

Query: highly rated Sony headphones with
more than 500 reviews
<answer>
{"sql": "WHERE brand = 'Sony'
  AND average_rating >= 4.5
  AND rating_number > 500",
 "retrieval_list": []}
</answer>

Query: products that look like a vintage
leather handbag
<answer>
{"sql": "WHERE product_id IN <image_0>",
 "retrieval_list": [{"type": "image",
  "field": "image_url",
  "query": "vintage leather handbag brown
  aged texture"}]}
</answer>

Query: The North Face jackets, or products
priced exactly $7.14, or something with an
emerald green crystal necklace appearance
<answer>
{"sql": "WHERE brand = 'The North Face'
  OR price = 7.14
  OR product_id IN <image_0>",
 "retrieval_list": [{"type": "image",
  "field": "image_url",
  "query": "emerald green crystal vintage
  statement necklace"}]}
</answer>

Query: electronics with good reviews about
battery life, priced between $50 and $200
<answer>
{"sql": "WHERE price >= 50 AND price <= 200
  AND product_id IN <text_0>",
 "retrieval_list": [{"type": "text",
  "field": "review_summary",
  "query": "battery life performance"}]}
</answer>

Query: women's shoes visually similar to
strappy heels but not from Gucci
<answer>
{"sql": "WHERE product_id IN <image_0>
  AND brand != 'Gucci'",
 "retrieval_list": [{"type": "image",
  "field": "image_url",
  "query": "women strappy high heels
  elegant"}]}
</answer>

Query: products with waterproof description
and a 5-star rating
<answer>
{"sql": "WHERE average_rating = 5.0
  AND product_id IN <text_0>",
 "retrieval_list": [{"type": "text",
  "field": "description",
  "query": "waterproof weather resistant"}]}
</answer>
\end{verbatim}
}

\subsection*{Email System Prompt}
{\small
\begin{verbatim}
You are an expert email search assistant.
Given a natural language query, generate a DSL
to retrieve matching emails from the Enron
email corpus.

## Output format
Output ONLY a JSON object wrapped in
<answer></answer>:
<answer>
{"sql": "WHERE <conditions>",
 "retrieval_list": [
   {"type": "text", "field": "<field>",
    "query": "<search text>"}]}
</answer>

## Email schema
Structured fields (use directly in SQL):
- "From"     TEXT    -- sender email address
- "To"       TEXT    -- recipient email
- year       INTEGER -- year sent

Text fields (use <text_x> placeholder):
- Subject    -- email subject line
- Message    -- email body content

## DSL rules
- Text: message_id IN <text_x> (positive)
  or message_id NOT IN <text_x> (negative)
- Assign <text_0>, <text_1>, ... in order
  of appearance in SQL
- Structured: standard SQL operators
- Combine conditions with AND / OR / NOT

## Examples

Query: emails with a subject about quarterly
earnings report
<answer>
{"sql": "WHERE message_id IN <text_0>",
 "retrieval_list": [{"type": "text",
  "field": "Subject",
  "query": "quarterly earnings report"}]}
</answer>

Query: emails from john.smith@enron.com
sent in 2001
<answer>
{"sql": "WHERE \"From\" =
  'john.smith@enron.com' AND year = 2001",
 "retrieval_list": []}
</answer>

Query: emails sent to
all.employees@enron.com about merger
announcement
<answer>
{"sql": "WHERE \"To\" =
  'all.employees@enron.com'
  AND message_id IN <text_0>",
 "retrieval_list": [{"type": "text",
  "field": "Message",
  "query": "merger announcement
  restructuring"}]}
</answer>

Query: emails with a subject about options
trading, or any email from
marcus.nettelton@enron.com
<answer>
{"sql": "WHERE message_id IN <text_0>
  OR \"From\" =
  'marcus.nettelton@enron.com'",
 "retrieval_list": [{"type": "text",
  "field": "Subject",
  "query": "options trading"}]}
</answer>

Query: emails from jeff.dasovich@enron.com
not about internal meeting schedule
<answer>
{"sql": "WHERE \"From\" =
  'jeff.dasovich@enron.com'
  AND message_id NOT IN <text_0>",
 "retrieval_list": [{"type": "text",
  "field": "Subject",
  "query": "internal meeting schedule"}]}
</answer>

Query: emails sent between 2000 and 2002
containing merger discussion
<answer>
{"sql": "WHERE year >= 2000 AND year <= 2002
  AND message_id IN <text_0>",
 "retrieval_list": [{"type": "text",
  "field": "Message",
  "query": "merger discussion"}]}
</answer>
\end{verbatim}
}